\documentclass[journal,12pt,draftclsnofoot,onecolumn]{IEEEtran}

\IEEEoverridecommandlockouts

\usepackage{amsthm}
\usepackage{amsfonts}
\usepackage{amssymb}
\usepackage{mathrsfs}
\usepackage{mathtools}
\usepackage[english]{babel}
\usepackage{float}
\usepackage{textcomp}

\usepackage[pdftex]{graphicx}
\usepackage{epstopdf}
\usepackage{amsmath}
\usepackage{xcolor}
\usepackage{url}
\usepackage{multirow}
\usepackage{graphicx}

\DeclareMathOperator*{\argmin}{arg\,min}
 
\usepackage{bm}
\usepackage{balance}
\usepackage{microtype}
\usepackage[caption=false,font=footnotesize]{subfig}
\usepackage{cite}
\usepackage{comment}
\allowdisplaybreaks

\newtheorem{theorem}{Theorem}
\makeatletter
\newcommand{\settheoremtag}[1]{
  \let\oldthetheorem\thetheorem
  \renewcommand{\thetheorem}{#1}
  \g@addto@macro\endtheorem{
    \addtocounter{theorem}{-1}
    \global\let\thetheorem\oldthetheorem}
  }
\makeatother

\usepackage[switch]{lineno}
\usepackage[named]{algo}
\usepackage[noend]{algpseudocode}
\usepackage{algorithm}

\DeclareMathOperator{\Tr}{tr}

\newcommand{\bsym}{\boldsymbol}

\newcommand{\alpharl}{\bsym{\alpha}_r^k[\ell]}
\newcommand{\taurl}{\bsym{\bar{\tau}}_r[\ell]}
\newcommand{\nurl}{[\bsym{\bar{\nu}}_r]_\ell}
\newcommand{\alphacq}{\bsym{\alpha}_c^k[q]}
\newcommand{\taucq}{\bsym{\bar{\tau}}_c[q]}
\newcommand{\nucq}{[\bsym{\bar{\nu}}_c]_q}

\begin{document}
%
\title{Factor Graph Processing for Dual-Blind Deconvolution at ISAC Receiver }

\author{Roman Jacome~\IEEEmembership{Student Member,~IEEE}, Edwin Vargas~\IEEEmembership{Student Member,~IEEE}, Kumar Vijay Mishra~\IEEEmembership{Senior Member,~IEEE}, Brian M. Sadler~\IEEEmembership{Life Fellow,~IEEE}, Henry Arguello~\IEEEmembership{Senior Member,~IEEE}
\thanks{R. J., E. V., and H. A. are with Universidad Industrial de Santander, Bucaramanga, Santander 680002 Colombia, e-mail: \{roman2162474@correo., edwin.vargas@correo., henarfu@\}uis.edu.co.}
\thanks{K. V. M. and B. M. S. are with the United States DEVCOM Army Research Laboratory, Adelphi, MD 20783 USA, e-mail: kvm@ieee.org, brian.m.sadler6.civ@mail.mil.}
\thanks{This research was sponsored by the Army Research Office/Laboratory under Grant Number W911NF-21-1-0099, the VIE project 9834 entitled ``Dual blind deconvolution for joint radar-communications processing'' and This work was supported by the VIE of the Universidad Industrial de Santander, Colombia under the research projects 3735. K. V. M. acknowledges partial support from the National Academies of Sciences, Engineering, and Medicine via the Army Research Laboratory Harry Diamond Distinguished Fellowship. The research was sponsored by the Army Research Laboratory and was accomplished under Cooperative Agreement Number W911NF-21-2-0288. The views and conclusions contained in this document are those of the authors and should not be interpreted as representing the official policies, either expressed or implied, of the Army Research Laboratory or the U.S. Government. The U.S. Government is authorized to reproduce and distribute reprints for Government purposes notwithstanding any copyright notation herein.}
}

\maketitle
\IEEEpeerreviewmaketitle

\begin{abstract}
Integrated sensing and communications (ISAC) systems have gained significant interest because of their ability to jointly and efficiently access, utilize, and manage the scarce electromagnetic spectrum. The co-existence approach toward ISAC focuses on the receiver processing of overlaid radar and communications signals coming from independent transmitters. A specific ISAC coexistence problem is \textit{dual-blind deconvolution} (DBD), wherein the transmit signals and channels of both radar and communications are unknown to the receiver. Prior DBD works ignore the evolution of the signal model over time. In this work, we consider a dynamic DBD scenario using a linear state space model (LSSM) such that, apart from the transmit signals and channels of both systems, the LSSM parameters are also unknown. We employ a \textit{factor graph} representation to model these unknown variables. We avoid the conventional matrix inversion approach to estimate the unknown variables by using an efficient expectation-maximization algorithm, where each iteration employs a Gaussian message passing over the factor graph structure. Numerical experiments demonstrate the accurate estimation of radar and communications channels, including in the presence of noise. 
\end{abstract}

\section{Introduction}
With the advent of mobile communications and new radar techniques, the already limited electromagnetic spectrum has become very inadequate for new applications \cite{mishra2019toward,elbir2023the}. In this context, joint radar-communications (ISAC) systems have recently gained prominence by offering various solutions to efficiently share the hardware and spectrum \cite{paul2016survey}. In general, ISAC systems are based on opportunistic \cite{duggal2020doppler}, coexistence \cite{ayyar2019robust}, or codesign \cite{liu2020co} methods. The co-existence ISAC approach is challenging because both radar and communications transmissions are independent or uncoordinated \cite{vargas2023dual}. The coexistence ISAC performance mainly relies on optimal receiver processing. In this paper, we focus on the coexistence scenario. 

The coexistence receiver admits overlaid radar and communication signals of which the radar transmit signal (communications channel) is known (estimated \textit{a priori}) and the radar channel (communications transmit signal) is unknown. However, when the radar is passive or multi-static \cite{wu2022resource}, its transmit signal may also be unknown to the receiver. Additionally, in wireless communications over dynamic channels like millimeter-wave \cite{dokhanchi2019mmwave} or terahertz-band \cite{elbir2021terahertz}, the channel coherence times are extremely brief \cite{song2011present}. Consequently, estimates of channel states become outdated rapidly. Therefore the communications channel becomes another unknown variable at the receiver. This leads to the \textit{dual-blind deconvolution} (DBD), where the transmit signals and channels of both radar and communications are unknown. 

Previous works have addressed the DBD problem by casting it as atomic norm minimization (ANM) \cite{vargas2023dual,jacome2023multi}. This approach harnesses the sparsity of the communications and radar channels and the low-dimensional subspace structure of the respective transmit signals to accurately estimate all the unknowns. However, ANM is solved through a computationally expensive semidefinite program. Some alternate approaches to mitigate the complexity include Beurling-Selberg extremization \cite{monsalve2022beurling}, which requires solving for low-rank Hankel-type matrix recovery. However, none of these approaches consider signal evolution over time. 

Contrary to prior DBD studies, we consider a dynamic signal setting modeled using a linear state-space model (LSSM), which includes hidden state variables. The LSSM has been employed earlier for both radar and communications. For radar signal processing, \cite{naishadham2008robust} used LSSM formulation to estimate the amplitude and phase of the extended scatterers. The sparse reconstruction processing of ultra-wideband radar in \cite{ren2018short} utilizes LSSM to analyze micro-Doppler signatures \cite{ren2018short}. Similarly, in wireless communications, \cite{vogt2019state} used LSSM to characterize the dynamics of a full-duplex communications channel. 
For single-blind deconvolution, \cite{zhang2000blind} employs LSSM in a sparse Bayesian learning framework. In particular, \cite{bruderer2015deconvolution} estimates a weakly sparse input signal without knowledge of the system parameters. It employs a distribution prior, usually a Gaussian, such that the sorted samples from its probability density function exhibit a power-law decay. Then, the expectation maximization (EM) \cite{wen2012data,hajek2015random} algorithm estimates the unknown variables. In \cite{zalmai2016blind}, EM is used to jointly learn the LSSM parameters and the sparse signal input in a BD problem. The EM iterations rely on building a probabilistic \textit{factor graph} \cite{loeliger2004introduction} and computing Gaussian message passing (GMP) \cite{dauwels2009expectation} to update each variable. These approaches have shown that estimating the unknown BD variables using EM is computationally efficient because GMP is computed using the modified Byron-Faizer smoother (MBFS) \cite{martin2013modified}, which is a stable method to propagate forward and backward messages in the factor graph structure.

In this work, to solve the LSSM-based DBD, we develop a maximum likelihood formulation, wherein Gaussian prior is used for the radar and communications channels to promote sparsity in the estimation process. We apply EM to learn the parameters of both LSSM and sparse channels. The EM updates require the computation of the expected value of several variables, which are efficiently derived using MBFS. Our numerical experiments show that the proposed approach yields high accuracy in the estimated parameters for varying levels of noise and number of observation samples.

The rest of the paper is organized as follows. In the next section, we present the LSSM signal model for the DBD receiver. Section \ref{sec:algorithm} presents the EM algorithm based on the probabilistic factor graph. We validate our model and methods through numerical experiments in Section \ref{sec:simulations} and conclude in \ref{sec:summary}.
\section{System Model}
\label{sec:model} 
 Consider a dynamic joint radar-communications system in a co-existence scenario. For notational simplicity, variables associated with radar (communications) are indicated with subscript $r$ ($c$). For instance, the radar transmit $K$ pulses, where each waveform is denoted by $\mathbf{f}_r(t)[k] \in \mathbb{C}^n$  and, the communications signal transmits $K$ messages $\mathbf{f}_c(t)[k] \in \mathbb{C}^n$, respectively. Both signals are bandlimited within $[-B/2, B/2]$. Consider the Swerling II model \cite{swerling1960probability} with $L$ fluctuating radar targets, the \textit{radar channel} for the transmit pulse $k$, is modeled as 
 $$h_r(t)[k] = \sum_{\ell=0}^{L-1}\alpharl \delta(t-\taurl) e^{-\mathrm{j}2\pi k\nurl},\; k=1,\dots,K,$$ 
 where $K$ is the number of pulses. Similarly, the communications channel models $Q$ fluctuating propagation paths, for the $k$ pulse  
 $$h_c(t)[k] = \sum_{q=0}^{Q-1}\alphacq  \delta(t-\taucq)e^{-\mathrm{j}2\pi k\nucq},\; k=1,\dots,K,$$ 
 where $\alphacq$ is the path loss varying in each pulse and $\taucq$ is the path delay. The overlaid received signal is 
 $$\mathbf{y}[k](t) = \mathbf{f}_r(t)[k]*h_r(t)[k] +\mathbf{f}_c(t)[k]*h_c(t)[k] + \mathbf{z}(t)[k],$$  where $\mathbf{z}[k] \sim \mathcal{N}(0,\mathbf{I}_n\sigma_z^2)$ is additive noise. 
 Consider the slow-time samples at a fixed range-time or delay and $k$-th pulse as 
 \begin{equation}
     \mathbf{y}[k] = \mathbf{f}_r[k]h_r[k] + \mathbf{f}_c(t)h_c[k]  + \mathbf{z}[k],\;\in\mathbb{C}^n,
 \end{equation}
 where $$h_r[k] = \sum_{\ell=0}^{L-1}\alpharl e^{-\mathrm{j}2\pi k\nurl},$$ and $$h_c[k] = \sum_{\ell=0}^{L-1}\alphacq e^{-\mathrm{j}2\pi k\nucq}.$$ 
 
 We aim to model the signal evolving through all the pulses with an LSSM \cite{prasanth2011state}. Define the state vector containing the $\ell$-th target response for the radar signal 
 \begin{align}
     &\mathbf{x}_{r\ell}[k] = \alpharl e^{-\mathrm{j}2\pi \nurl} \mathbf{x}_{r\ell}[k-1]\\
     &\mathbf{y}_{r\ell}[k] = \mathbf{f}_r[k]\mathbf{x}_{r\ell}[k-1],
 \end{align}
 where $\mathbf{x}_{r\ell}[k] \in \mathbb{C}^n$. Then, the state vector for the $q$-th path of the communications signal as 
 \begin{align}
     &\mathbf{x}_{cq}[k] = \alpharl e^{-\mathrm{j}2\pi \nucq} \mathbf{x}_{cq}[k-1]\\
     &\mathbf{y}_{q\ell}[k] = \mathbf{f}_q[k]\mathbf{x}_{cq}[k-1].
 \end{align}
The LSSM for the overlaid receiver is 
\begin{align}
     &\mathbf{x}_r[k] = \mathbf{A}_r\mathbf{x}_r[k-1] + {\mathbf{f}}_r[k]h_r[k] + \mathbf{e}_r[k]\nonumber\\&\mathbf{x}_c[k] = \mathbf{A}_c\mathbf{x}_c[k-1] + {\mathbf{f}}_c[k]h_c[k] + \mathbf{e}_c[k]\nonumber\\&y[k] =\mathbf{x}_r[k] + \mathbf{x}_c[k] + \mathbf{z}[k],\label{lssm}
 \end{align}
where  $\mathbf{A}_r\in\mathbb{C}^{n\times n}$ and $\mathbf{A}_c\in\mathbb{C}^{n\times n}$ are the corresponding state matrices; $\mathbf{e}_r\in \mathbb{C}^{n}$, $\mathbf{e}_c\in \mathbb{C}^{n}$ are random noise vectors during the transmission. Denote the  channel vectors as $\mathbf{h}_r = [h_r[1],\dots,h_r[k]] \in \mathbb{C}^{K} $ and  $\mathbf{h}_c = [h_c[1],\dots,h_c[k]] \in \mathbb{C}^{K}$.
 
\begin{figure}[t!]
    \centering
\includegraphics[width=0.7\linewidth]{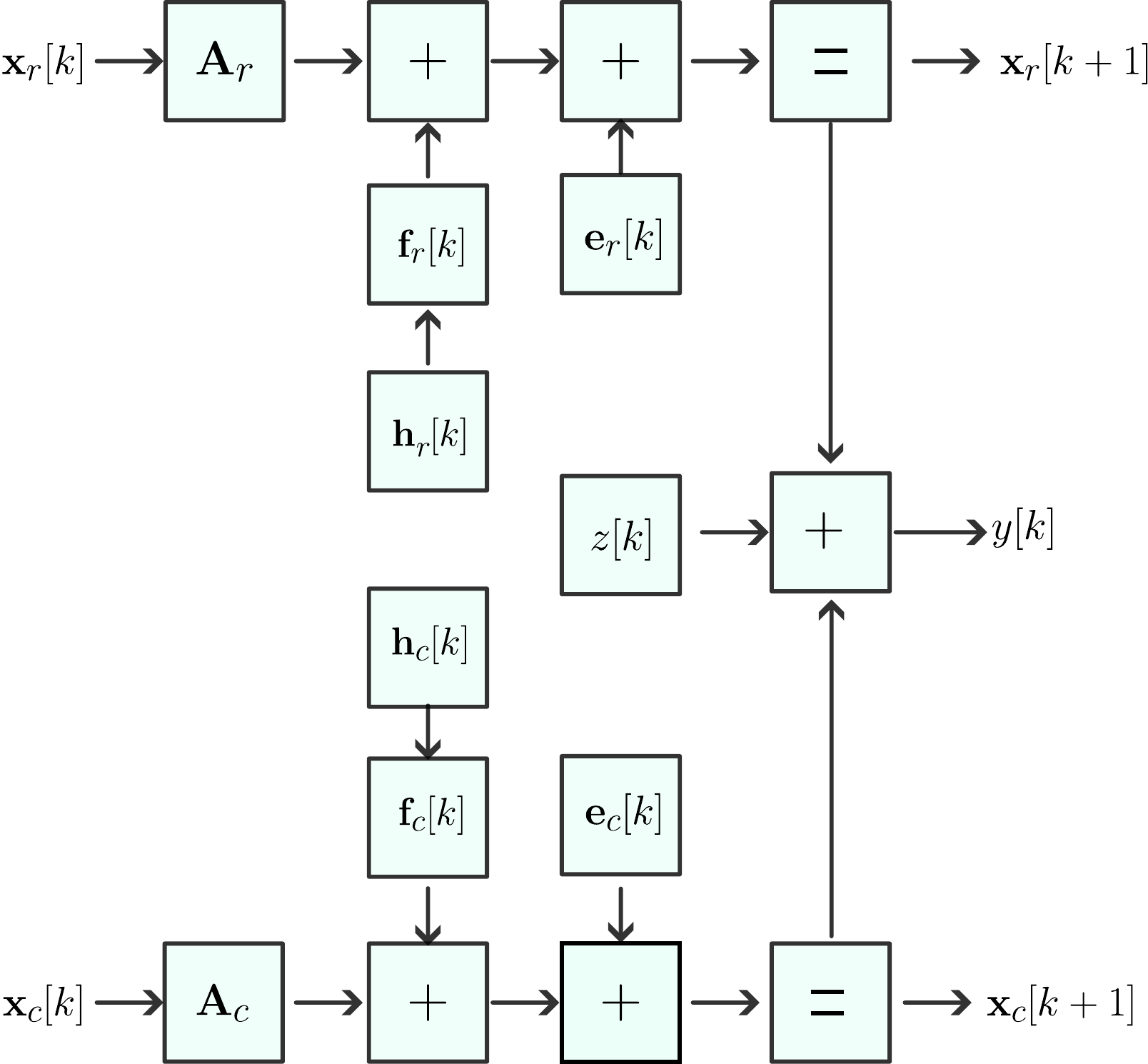}
    \caption{Factor graph representation of the LSSM for dynamic ISAC receiver \eqref{lssm}, where the boxes include the unknown variables and factors. 
    }
    \label{fig:graph}
\end{figure}

Our goal is to estimate the sparse channel vectors $\mathbf{h}_r$ and  $\mathbf{h}_c$ related to the channels from the measurements $\mathbf{y}$ without the knowledge of the transmit signals ${\mathbf{f}}_r[k]$ and ${\mathbf{f}}_c[k]$ along with the model parameters $\mathbf{A}_r$, and $\mathbf{A}_c$, under the noise perturbation variables $\mathbf{e}_r[k] \sim \mathcal{N}(0,\sigma_{e_{r}}^2)$ and $\mathbf{e}_c[k] \sim \mathcal{N}(0,\sigma_{e_{c}}^2)$ and $z[k] \sim \mathcal{N}(0,\sigma_z^2)$. To this end, we employ a factor graph (Fig. \ref{fig:graph}) to describe the algorithm that estimates the unknown variables \cite{loeliger2004introduction}. 
The nodes/boxes in the factor graph denote both unknown variables and factors (operations between the variables). Based on this structure, we employ a GMP along the graph to perform an EM algorithm to estimate the unknown variables. The GMP approach employed is based on the MBFS method to compute the expectation of the random variables along the graph.
\section{DBD EM Method}\label{sec:algorithm}
 
The proposed algorithm employs the sparsity of channels radar and communications as $||\mathbf{h}_r||_0 = L\ll K$ and $||\mathbf{h}_c||_0 = Q\ll K$. Consider the random vectors $\mathbf{v}_c=[{v}_c[1], \cdots,{v}_c[K]\}$, $\mathbf{v}_r=[{v}_r[1], \cdots,{v}_r[K]]$ drawn from a Gaussian distribution with variances denoted by $\sigma_{\mathbf{v}_c}^2=[\sigma_{\mathbf{v}_c[0]}^2,\cdots,\sigma_{\mathbf{v}_c[K]}^2]$ and $\sigma_{\mathbf{v}_r}^2=[\sigma_{\mathbf{v}_r[0]}^2,\cdots,\sigma_{\mathbf{v}_r[k]}^2]$ such that $\mathbf{v}_c[k] \sim \mathcal{N}(0,\sigma_{\mathbf{v}_c[k]}^2)$,  $\mathbf{v}_r[k] \sim \mathcal{N}(0,\sigma_{\mathbf{v}_r[k]}^2)$ for {$k = 1,\dots,K$}. We aim to recover the system channels $\mathbf{h}_r$ and $\mathbf{h}_c$ up to a trivial ambiguity. The ambiguity arises from the products with the unknown input signals ${\mathbf{f}}_r[k]$ and ${\mathbf{f}}_c[k]$, i.e, for a non-zero scalar $\beta$ we have that $(\beta{\mathbf{f}}_r[k]) \mathbf{h}_r[k] = {\mathbf{f}}_r[k](\mathbf{h}_r[k]\beta)$ which is a usual scenario in inverse problems such as in blind deconvolution \cite{abed1997blind}. 
  
\subsection{Maximum Likelihood Formulation}
  
Define the discretized measurement vector by collecting the measurements as $\mathbf{y} = [y[1],\dots,y[k]]$. We choose a proper prior for $\mathbf{h}_r$ and $\mathbf{h}_c$ by, respectively, employing the conditional distributions
\begin{equation}\label{eq:prior_c}
 p(\mathbf{h}_c|\sigma_{\mathbf{v}_c}^2)=\prod_{k=1}^K \frac{1}{\sqrt{2\pi\sigma_{\mathbf{v}_c[k]}^2}}\exp\left(-\frac{\left(h_c[k]\right)^2}{2\sigma_{\mathbf{v}_c[k]}^2}\right),
       \end{equation}
       and
       \begin{equation}\label{eq:prior_r}
        p(\mathbf{h}_r|\sigma_{\mathbf{v}_r}^2)=\prod_{k=1}^K \frac{1}{\sqrt{2\pi\sigma_{\mathbf{v}_r[k]}^2}}\exp\left(-\frac{\left(h_r[k]\right)^2}{2\sigma_{\mathbf{v}_r[k]}^2}\right).
\end{equation}

Define the likelihood functions of radar and communications state vectors $\mathbf{x}_r$ and $\mathbf{x}_c$ as, respectively,
\begin{equation}\label{eq:likelihood_r}
    \begin{split}
        &p(\mathbf{y}|\mathbf{\theta},\mathbf{h}_r)=\int  \prod_{k=1}^K \frac{1}{\sqrt{2\pi \sigma_z^2}}e^{-\frac{1}{2\sigma_z^2}\left({y}[k]- \mathbf{x}_c[k]- \mathbf{x}_r[k]\right)^2} \\
        &\cdot\frac{1}{\left(2\pi \sigma_{{e_r}}^2 \right)^{\frac{n}{2}}}e^{-\frac{1}{2\sigma_{{e_r}}^2}\left|\left|\mathbf{x}_r[k]-\left(\mathbf{A}_r \mathbf{x}_r[k-1] + {\mathbf{f}}_r[k]h_r[k]\right)\right|\right|_2^2}d\mathbf{x}_r,
    \end{split}
\end{equation}
and
\begin{equation}\label{eq:likelihood_c}
    \begin{split}
        &p(\mathbf{y}|\mathbf{\theta},\mathbf{h}_c)=\int \prod_{k=1}^K \frac{1}{\sqrt{2\pi \sigma_z^2}}e^{-\frac{1}{2\sigma_z^2}\left({y}[k]- \mathbf{x}_c[k]- \mathbf{x}_r[k]\right)^2} \\
        &\cdot\frac{1}{\left(2\pi \sigma_{\mathbf{e_c}}^2 \right)^{\frac{n}{2}}}e^{-\frac{1}{2\sigma_{{e_c}}^2}\left|\left|\mathbf{x}_c[k]-\left(\mathbf{A}_c \mathbf{x}_c[k-1] + {\mathbf{f}}_c[k]h_c[k]\right)\right|\right|_2^2}d\mathbf{x}_c.
    \end{split}
\end{equation}
 With the prior distributions \eqref{eq:prior_c}-\eqref{eq:prior_r} and the likelihood functions \eqref{eq:likelihood_c}-\eqref{eq:likelihood_r}, we estimate the unknown variables denoted by  $\bsym{\theta}=\{\mathbf{A}_c,\mathbf{A}_r, $ $, ,\mathbf{f}_c[1],\dots,\mathbf{f}_c[k],\mathbf{f}_r[1],\dots,\mathbf{f}_r[k],\sigma_z^2,\sigma_{\mathbf{v}_c}^2,\sigma_{\mathbf{v}_r}^2\}$ by maximizing the marginal likelihood
\begin{align}\label{likelihood}
       L(\bsym{\theta})=p(\mathbf{y}|\bsym{\theta})=\iint &p(\mathbf{y}|\mathbf{\theta},\mathbf{h}_c)p(\mathbf{y}|\mathbf{\theta},\mathbf{h}_r)p(\mathbf{h}_c|\sigma_{\mathbf{v}_c}^2)\nonumber\\&p(\mathbf{h}_r|\sigma_{\mathbf{v}_r}^2)d\mathbf{h}_c d\mathbf{h}_r,
\end{align}
Note that estimating the variables $\sigma_{\mathbf{v}_r}^2$ and $\sigma_{\mathbf{v}_c}^2$ implies estimating $\mathbf{h}_r$ and $\mathbf{h}_c$, respectively.
 
\subsection{Expectation Maximization Algorithm}
 
The EM algorithm updates the parameters according to 
\begin{equation}\label{eq:optimization}
    \widehat{\bsym{\theta}}= \underset{\bsym{\theta}}{\operatorname{argmax}} \mathbb{E}\left[\ln p(\mathbf{y},\mathbf{v}_c,\mathbf{v}_r,{\mathbf{x}}_c,\mathbf{x}_r|\mathbf{\theta})  \right],
\end{equation}
where the joint density function $\ln p(\mathbf{y},\mathbf{h}_c,\mathbf{h}_r,\mathbf{x}_c,\mathbf{x}_r|\mathbf{\theta})$ is 
\begin{align}\label{eq:joint}
        -2\ln p(\mathbf{y},\mathbf{h}_c,\mathbf{h}_r,\mathbf{x}_c,\mathbf{x}_r|\bsym{\theta})&=-2\ln \left( \frac{\partial p(\mathbf{y}|\mathbf{\theta},\mathbf{h}_c,\mathbf{h}_r)}{\partial{\mathbf{x}}_c \partial{\mathbf{x}}_r} \right) \nonumber\\ 
        &= \sum_{k=1}^K \left( \frac{\left(y[k]- \mathbf{x}_c[k]- \mathbf{x}_r[k]\right)}{\sigma_z^2}\right.+\left. \frac{\left|\left|\mathbf{x}_c[k]-\left(\mathbf{A}_c \mathbf{x}_c[k-1] + {\mathbf{f}}_c[k]h_c[k]\right)\right|\right|_2^2}{\sigma_{{e_c}}^2}\right.\nonumber\\
        &+ \left.\frac{\left|\left|\mathbf{x}_r[k]-\left(\mathbf{A}_r \mathbf{x}_r[k-1] + {\mathbf{f}}_r[k]h_r[k]\right)\right|\right|_2^2}{\sigma_{{e_r}}^2}\right.\left.+ \ln \left(2\pi \sigma_z^2 \left(2\pi \sigma_{{e_c}}^2\sigma_{{e_r}}^2 \right)^n  \right) \right).
\end{align}
Then, the parameter update is obtained by 
\begin{equation}
    \widehat{\bsym\theta} = \argmin_{\bsym{\theta}} -\mathbb{E}\left[\ln p(\mathbf{y},\mathbf{h}_c,\mathbf{h}_r,\mathbf{x}_c,\mathbf{x}_r|\bsym{\theta})\right].\label{eq:problem_opt}
\end{equation}
Using the expression above, we derive closed-form solutions for each parameter in $\bsym\theta$. We update the variances $\sigma_{\mathbf{v}_r}$ and $\sigma_{\mathbf{v}_c}$ by first taking the derivatives of \eqref{eq:joint} with respect to $\sigma_{\mathbf{v}_r}$ and $\sigma_{\mathbf{v}_c}$. Equating the derivatives to zero yields
\begin{align}\label{eq:first}
       \widehat{ \sigma}_{\mathbf{v}_c[k]}^2&=\underset{\mathbf{\sigma_{\mathbf{v}_c[k]}^2}}{\operatorname{argmin}} \frac{1}{\sigma_{\mathbf{v}_c[k]}^2}\mathbb{E}\left[\left(\mathbf{v}_c[k]\right)^2\right] + \ln \left( 2\pi \mathbf{v}_c[k] \right) =\mathbb{E}\left[\left(\mathbf{v}_c[k]\right)^2\right],
       \end{align}
       \begin{align}
       \widehat{ \sigma}_{\mathbf{v}_r[k]}^2&=\underset{\mathbf{\sigma_{\mathbf{v}_r[k]}^2}}{\operatorname{argmin}} \frac{1}{\sigma_{\mathbf{v}_r[k]}^2}\mathbb{E}\left[\left(\mathbf{v}_r[k]\right)^2\right] + \ln \left( 2\pi \mathbf{v}_r[k] \right) =\mathbb{E}\left[\left(\mathbf{v}_r[k]\right)^2\right].
\end{align}

Similarly, taking derivatives of \eqref{eq:joint} with respect to $\mathbf{f}_r[k]$ and $\mathbf{f}_c[k]$ for $k=1,\dots, K$ and equating them to zero provides the estimates of the transmit signals as
\begin{align}\label{eq14}
\widehat{{\mathbf{f}}}_c[k]=\frac{\mathbb{E}\left[\mathbf{v}_c[k]\mathbf{x}_c[k]\right] - \widehat{\mathbf{A}}_c\mathbb{E}\left[\mathbf{v}_c[k]\mathbf{x}_c[k-1]\right]}{\mathbb{E}\left[\left(\mathbf{v}_c[k]\right)^2\right]}, \\
        \widehat{{\mathbf{f}}}_r[k]=\frac{\mathbb{E}\left[\mathbf{v}_r[k]\mathbf{x}_r[k]\right] - \widehat{\mathbf{A}}_r\mathbb{E}\left[\mathbf{v}_r[k]\mathbf{x}_r[k-1]\right]}{\mathbb{E}\left[\left(\mathbf{v}_r[k]\right)^2\right]}.
\end{align}
Finally, to estimate the state matrices $\mathbf{A}_r$ and $\mathbf{A}_c$, we solve the following quadratic optimization problems:
\begin{equation}\label{eq:a_c}
        \widehat{\mathbf{A}}_c= \underset{\mathbf{A}_c}{\operatorname{argmin}} \Tr  \left(\mathbf{A}_c \mathbf{V}_{\mathbf{A}_c}\left(\mathbf{A}_c\right)^T -2\mathbf{A}_c\bsym{\Lambda}_{\mathbf{A}_c} \right), 
        \end{equation}
\begin{equation}\label{eq:a_r}
       \widehat{\mathbf{A}}_r= \underset{\mathbf{A}_r}{\operatorname{argmin}} \Tr  \left(\mathbf{A}_r \mathbf{V}_{\mathbf{A}_r}\left(\mathbf{A}_r\right)^T -2\mathbf{A}_r\bsym{\Lambda}_{\mathbf{A}_r} \right),
\end{equation}
where 
\begin{align}\label{eq17}
        \mathbf{V}_{\mathbf{A}_c}=&\sum_{k=1}^K \mathbb{E}\left[\mathbf{x}_c[k-1]\left(\mathbf{x}_c[k-1]\right)^T \right] - \frac{\mathbb{E}\left[\mathbf{v}_c[k]\mathbf{x}_c[k-1] \right]\mathbb{E}\left[\mathbf{v}_c[k]\mathbf{x}_c[k-1] \right]^T}{\mathbb{E}\left[\left(\mathbf{v}_c[k] \right)^2\right]}
        \end{align}
        \begin{align}
        \mathbf{V}_{\mathbf{A}_r}=&\sum_{k=1}^K \mathbb{E}\left[\mathbf{x}_r[k-1]\left(\mathbf{x}_r[k-1]\right)^T \right] - \frac{\mathbb{E}\left[\mathbf{v}_r[k]\mathbf{x}_r[k-1] \right]\mathbb{E}\left[\mathbf{v}_r[k]\mathbf{x}_r[k-1] \right]^T}{\mathbb{E}\left[\left( \mathbf{v}_r[k] \right)^2\right]}
\end{align}
and 
\begin{align} \label{eq18}
        \bsym{\Lambda}_{\mathbf{A}_c} &= \sum_{k=1}^K \mathbb{E}\left[\mathbf{x}_c[k-1]\left(\mathbf{x}_c[k]\right)^T \right] - \frac{\mathbb{E}\left[\mathbf{v}_c[k]\mathbf{x}_c[k-1] \right]\mathbb{E}\left[\mathbf{v}_c[k]\mathbf{x}_c[k] \right]^T}{\mathbb{E}\left[\left( \mathbf{v}_c[k] \right)^2\right]}
        \end{align}\vspace{-0.5cm}
        \begin{align}
        \bsym{\Lambda}_{\mathbf{A}_r} &= \sum_{k=1}^K \mathbb{E}\left[\mathbf{x}_r[k-1]\left(\mathbf{x}_r[k]\right)^T \right] - \frac{\mathbb{E}\left[\mathbf{v}_r[k]\mathbf{x}_r[k-1] \right]\mathbb{E}\left[\mathbf{v}_r[k]\mathbf{x}_r[k] \right]^T}{\mathbb{E}\left[\left( \mathbf{v}_r[k] \right)^2\right]}.\label{eq:last}
\end{align}
The quadratic problems in \eqref{eq:a_c}  and \eqref{eq:a_r} have closed-form solutions \cite{zalmai2017state} depending on the structure of the matrix. Assume both matrices are controllable, i.e. the matrices have ones in the first upper diagonal and the last row of the matrices contains the coefficients $\mathbf{a}_r\in\mathbb{R}^{n}$ and $\mathbf{a}_c\in\mathbb{R}^{n}$. We follow the formulation in \cite[Appendix D]{zalmai2017state} to solve \eqref{eq:a_c}. and \eqref{eq:a_r}. The initialization of the LSSM parameters $\widehat{\mathbf{A}}_r,\widehat{\mathbf{A}}_c$ and the unknown input signals ${\mathbf{f}}_r[k]$, ${\mathbf{f}}_c[k]$ are set to zeros as suggested in prior works \cite{zalmai2016blind,zalmai2017state}. We perform steps \eqref{eq:first} to \eqref{eq:last} for several iterations to increase the accuracy of the estimates.

We compute the expected values using GMP through the factor graph in Fig. \ref{fig:graph}. Particularly, these were computed using the modified Bryson-Frazier smoother \cite{gibbs2010square} which provides a highly efficient form to estimate these quantities. 
\begin{figure}[t]
    \centering
    \includegraphics[width=0.75\linewidth]{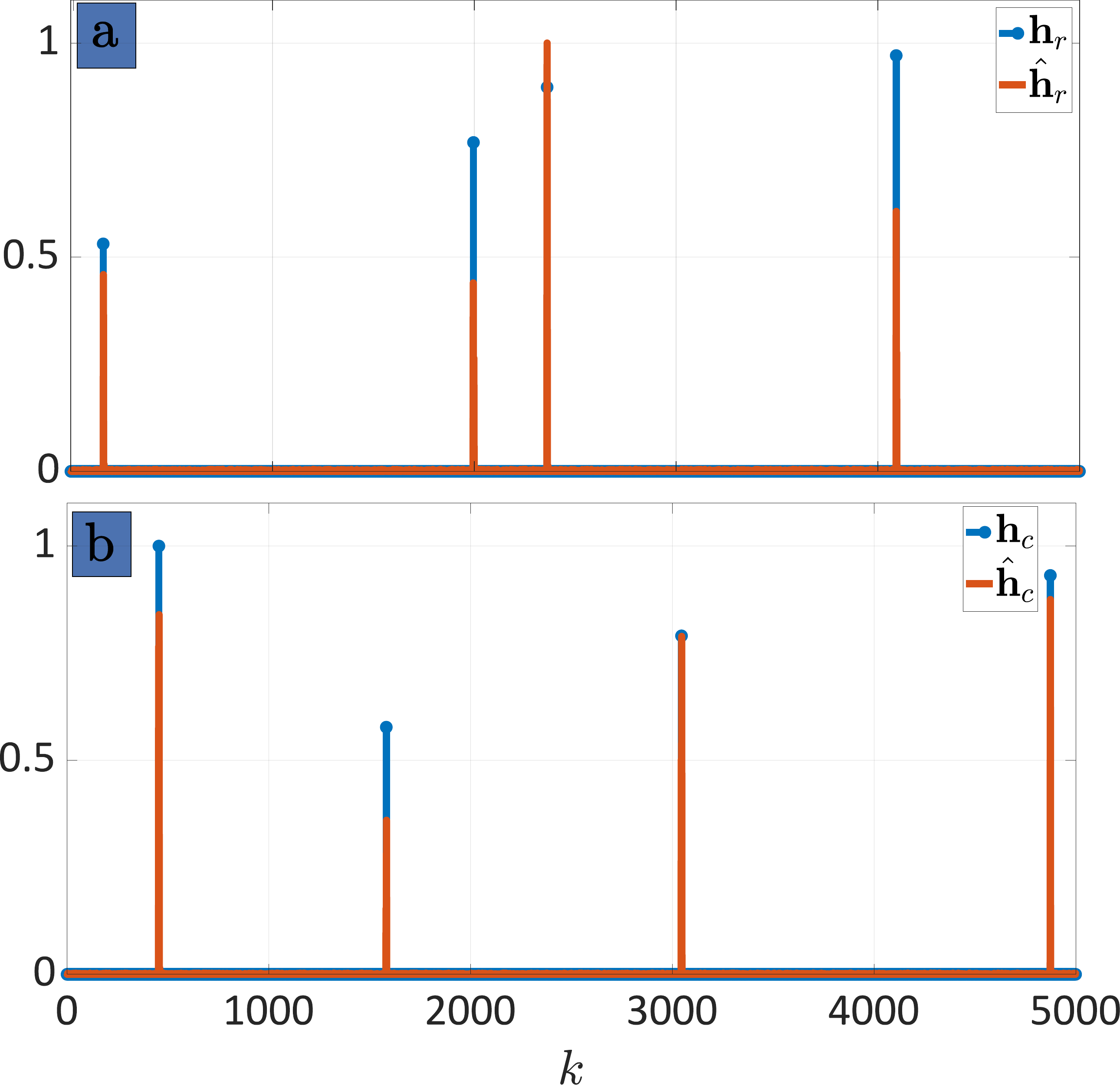}\vspace{-0.45cm}
    \caption{EM-based parameter estimation of (a) radar and (b) communications channels in the absence of noise. The plots show the time-indexed locations of spikes in the estimated (red) and true (blue) channels.  } 
    \label{fig:noiseless}
\end{figure}
\begin{figure}[t]
    \centering
    \includegraphics[width=0.75\linewidth]{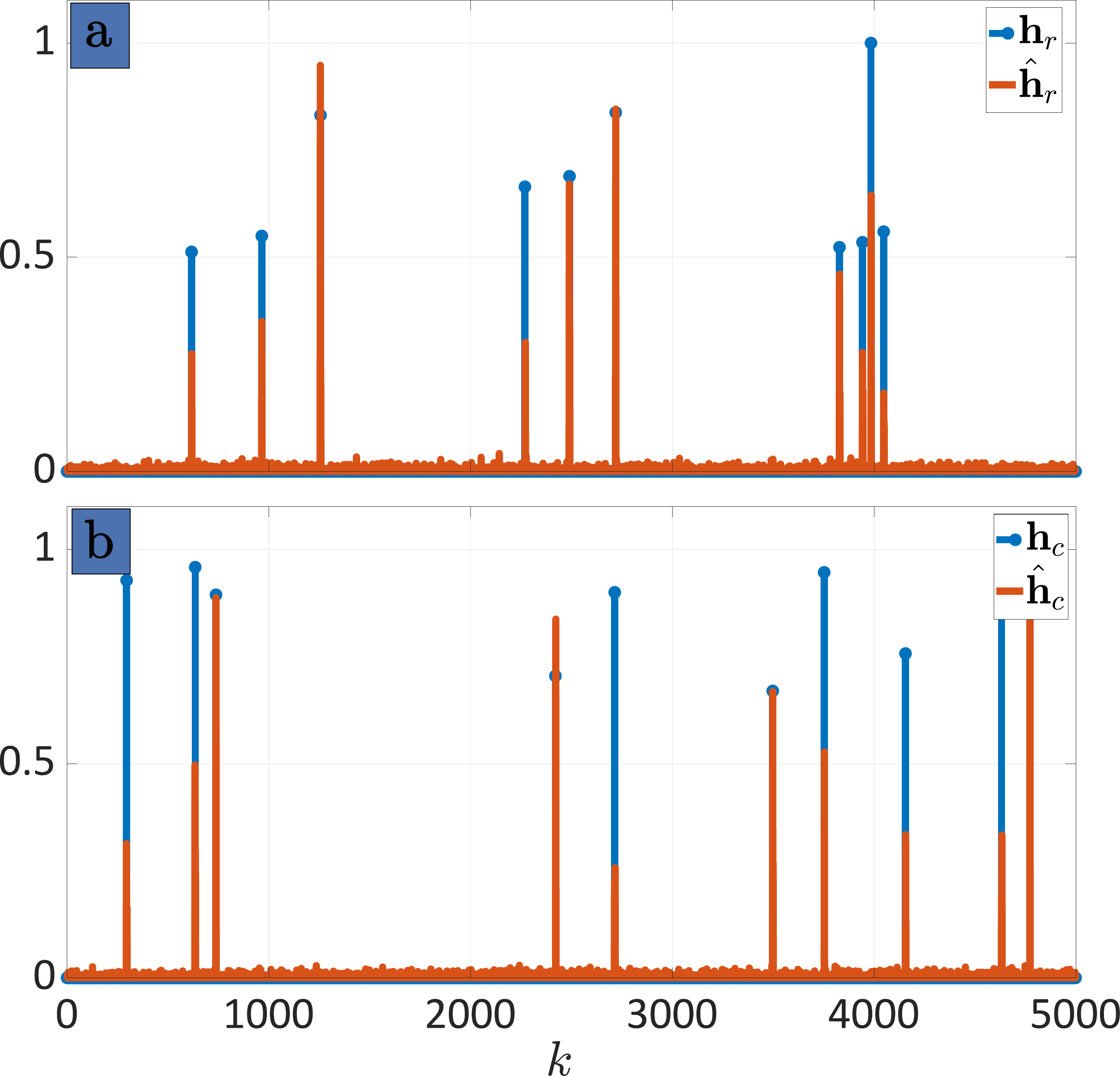}
    \caption{As in Fig.~\ref{fig:noiseless}, but in the presence of noise.  } 
    \label{fig:nosy}
\end{figure}
\begin{figure}[!t]
    \centering
    \includegraphics[width=0.75\linewidth]{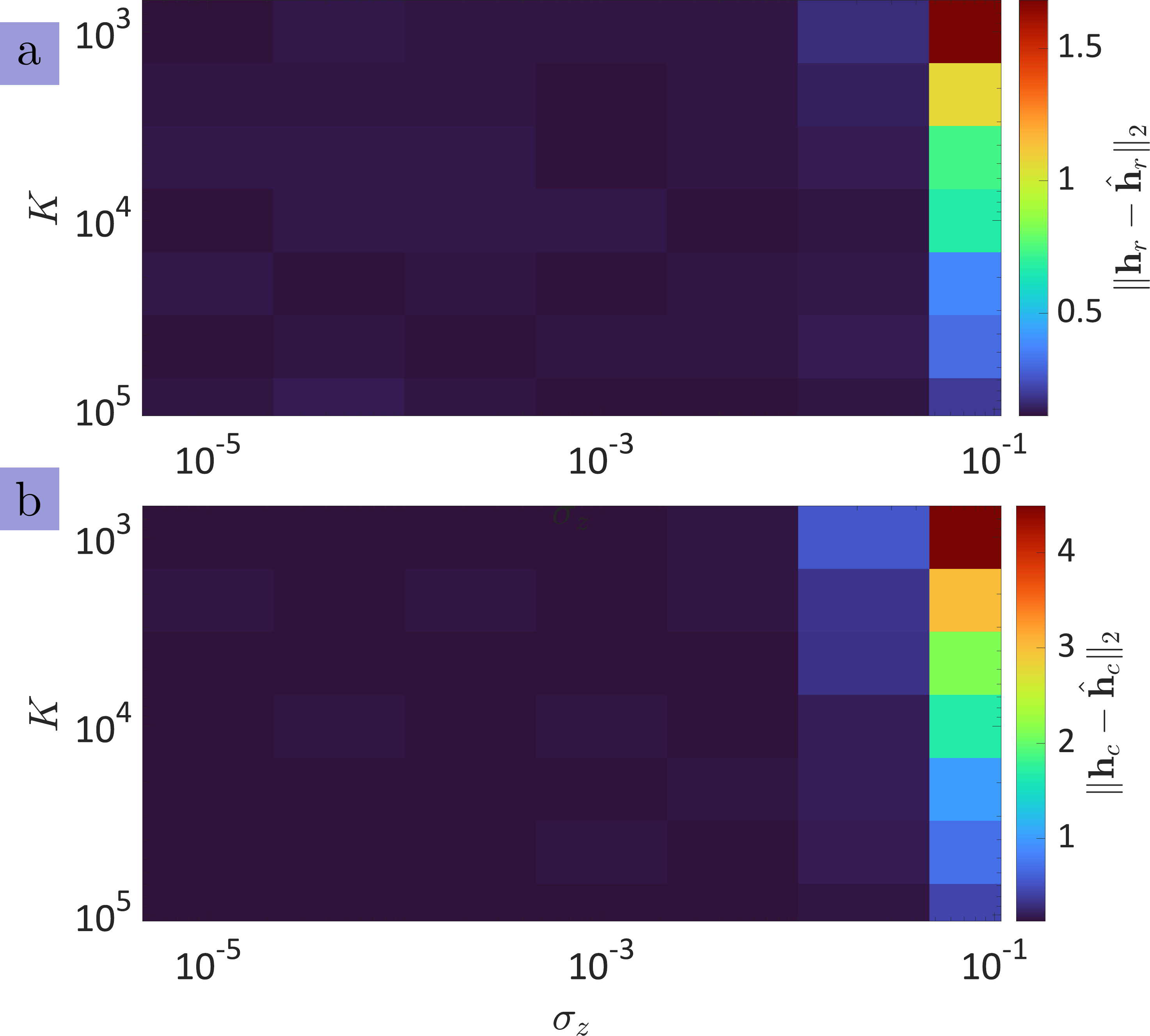}
    \caption{Channel estimation error for (a) radar and (b) communications for different values of $K$ and noise variance $\sigma_z^2$. Here, $n=4$ and $L=Q=3$.  }
    \label{fig:varaitions}
\end{figure}
 
\section{Numerical Experiments}
\label{sec:simulations}
 
We validated the proposed method through numerical experiments. We set the number of samples $K=5\times10^{3}$, the number of targets $L$, propagation paths $Q=L=4$, and the dimension $n=4$. The transmit signals were drawn from a random distribution i.e $\mathbf{f}_r[k]\sim \mathcal{N}(0,\mathbf{I}_n)$, $\mathbf{f}_c[k]\sim \mathcal{N}(0,\mathbf{I}_n)$. The locations of the parameters $\nurl$ and $\nucq$ were set randomly. State matrices $\mathbf{A}_r$ and $\mathbf{A}_c$ were constructed with a controllable structure, where their coefficients were drawn from the uniform distribution $\mathcal{U}(-1,1)$. The number of iterations was set to 100. 

We evaluated the performance of the algorithm for the noiseless case, i.e. $\sigma_z^2 = 0$. Fig. \ref{fig:noiseless} shows the estimation of the radar and communications channels, wherein the locations of the spikes refer to the delay parameters of the targets and the propagation paths, respectively. We show the exact recovery of the locations of the Doppler parameter and the amplitude difference is related to the trivial ambiguity between the channel and the transmit signals. Next, to validate the algorithm under a more challenging setting, we set $\sigma_z^2 = 10^{-2}$ and the channels' sparsity $L=Q=10$. Fig. \ref{fig:nosy} shows that, while the channel vector estimates have some small perturbation, the spikes of the channel parameters are accurately estimated. 

Finally, we performed Monte Carlo experiments by varying the number of samples $K$ from $10^{3}$ to $10^{5}$ and $\sigma_z^2$ from $10^{-3}$ to $10^{-1}$. In each case, we selected 7 values on a logarithmic scale and conducted 50 trials for each setting. For this experiment, we set $L=Q=3$ and $n=4$. We compute the Euclidean norm of the estimation errors, i.e., $\Vert\mathbf{h}_r -\widehat{\mathbf{h}}_r\Vert_2$ and $\Vert\mathbf{h}_c -\widehat{\mathbf{h}}_c\Vert_2$ for radar and communications, respectively. Fig. \ref{fig:varaitions} plots these errors averaged over all $50$ trials. The results suggest near-zero estimation error for almost all noise levels. The error expectedly increases with a decrease (increase) in the number of pulses (noise levels).

\section{Summary}\label{sec:summary}
 
We proposed a graph processing method for the dual-blind deconvolution recovery in a dynamic ISAC system. We formulated an LSSM for this scenario and derived the corresponding probabilistic factor graph. To estimate the channel vectors, we employed a sparsity-promoting EM algorithm following the GMP on the factor graph. Future investigations may include jointly estimating both time-delay and Doppler parameters \cite{vargas2022joint,vargas2023dual} and the case of multi-antenna receiver \cite{jacome2022multid,jacome2023multi}. 

\bibliographystyle{IEEEtran}
\bibliography{biblio.bib}
\end{document}